\def\sorb{b}
\newlength{\absize}

\if l\sorb \documentstyle{article}
\renewcommand{\baselinestretch}{1.5}
\renewcommand{\arraystretch}{1.5}
\begin{document}
\date{}
\pagestyle{empty}
\thispagestyle{empty}
\renewcommand{\thefootnote}{\fnsymbol{footnote}}
\newcommand{\starttext}{\newpage\normalsize
\pagestyle{plain}
\setlength{\baselineskip}{4ex}\par
\twocolumn\setcounter{footnote}{0}
\renewcommand{\thefootnote}{\arabic{footnote}}
}
\else
\documentstyle[12pt]{article}
\setlength{\absize}{6in}
\setlength{\topmargin}{-.5in}
\setlength{\oddsidemargin}{-.3in}
\setlength{\evensidemargin}{-.3in}
\setlength{\textheight}{9in}
\setlength{\textwidth}{7in}
\renewcommand{\baselinestretch}{1.5}
\renewcommand{\arraystretch}{1.5}
\setlength{\footnotesep}{\baselinestretch\baselineskip}
\begin{document}
\thispagestyle{empty}
\pagestyle{empty}
\renewcommand{\thefootnote}{\fnsymbol{footnote}}
\newcommand{\starttext}{\newpage\normalsize
\pagestyle{plain}
\setlength{\baselineskip}{4ex}\par
\setcounter{footnote}{0}
\renewcommand{\thefootnote}{\arabic{footnote}}
}
\fi

\input prepictex
\input pictex
\input postpictex
\newdimen\tdim
\tdim=\unitlength
\def\stpltsmbl{\setplotsymbol ({\small .})}
\def\tarrow{\arrow <5\tdim> [.3,.6]}
\def\figsize{}

\setcounter{bottomnumber}{2}
\setcounter{topnumber}{3}
\setcounter{totalnumber}{4}
\renewcommand{\bottomfraction}{1}
\renewcommand{\topfraction}{1}
\renewcommand{\textfraction}{0}

\newcommand{\preprint}[1]{\begin{flushright}
\setlength{\baselineskip}{3ex}#1\end{flushright}}
\renewcommand{\title}[1]{\begin{center}\LARGE
#1\end{center}\par}
\renewcommand{\author}[1]{\vspace{2ex}{\Large\begin{center}
\setlength{\baselineskip}{3ex}#1\par\end{center}}}
\renewcommand{\thanks}[1]{\footnote{#1}}
\renewcommand{\abstract}[1]{\vspace{2ex}\normalsize\begin{center}
\centerline{\bf Abstract}\par\vspace{2ex}\parbox{\absize}{#1
\setlength{\baselineskip}{2.5ex}\par}
\end{center}}

\def\theequation{\thesection.\arabic{equation}}
\preprint{\#HUTP-95/A033\\ 10/95}
\title{
A Perturbative Expansion for Weakly Bound States\thanks{Research
supported in part by the National Science Foundation under Grant
\#PHY-9218167.}
}
\author{
Hael Collins, Howard Georgi and David Zeltser\\
Lyman Laboratory of Physics \\
Harvard University \\
Cambridge, MA 02138 \\
}
\date{}
\abstract{We describe a perturbation expansion for the energy and wave
function of a weakly bound particle in a short-range potential in one space
dimension.
}
\starttext

\section{Introduction}

In this article, we discuss a method for calculating energies and wave
functions for weakly bound particles in short range potentials in one
dimensional nonrelativistic quantum mechanics. We usually think of the
formation of a bound state as a non-perturbative process, however, we will see
that for a weakly bound state, with a wave function that extends far outside
the region in which the potential is important, we can calculate bound state
properties in perturbation theory. In one dimension, the perturbation
expansion completely determines the energy and wave function of the state.
The expansion is quite simple (closely related to the Born series) and it may
well appear somewhere in the literature, but we have been unable to find it.

A particle moving in a generic short range potential has two natural
length scales: $a$, the range of the potential, and $\kappa^{-1}$, the size of
the state, related to the binding energy, $E_0$, of the system by
$\kappa=\sqrt{-2mE_0}$.\footnote{Throughout this article, we use
units in which $\hbar =1$.}  In turn, these scales define a
dimensionless parameter, $\kappa a$.  A weakly bound state, one for
which $\kappa a \ll 1$, inherently has a small dimensionless parameter
which could serve as the basis for a perturbative description of the
wave function.  In the following, we show that in one dimension we can
construct a power series in $\kappa a$ for both the energy and (in a sense)
the wave
function, assuming that $\kappa a \ll 1$.\footnote{The situation in three
dimensions is much more complicated. This condition is
replaced by an integral condition on the potential---the small
parameter then is the deviation of the actual potential from
this condition.}

The perturbative expansion that emerges is a series of $n$ coupled
integrals with $n$ powers of the potential and $n-1$ coupling terms at
each order of the power series.  These terms manifestly respect
translation invariance---if the potential is translated in space,
the wave function moves correspondingly.  For simple potentials, the
analytic behavior of the expansion for the bound states can easily be
extracted to several orders. We illustrate the procedure for a classic
square well potential in order to demonstrate the agreement of the
perturbative approach with the known form of the ground state.  Less
simple potentials can be handled numerically.

Our goal is eventually to extend this perturbative approach to three
dimensions and from there to fully relativistic systems.  The
extension of these one dimensional results to three is not as
straightforward as might be hoped.  The particle's wave function
varies significantly where the potential is large, making the
unraveling of the leading $\kappa a$ behavior considerably more
difficult.  Moreover, the form of the free particle Green's function
in three dimensions, unlike its one dimensional analogue, is singular
in its first term when expanded in powers of $\kappa a$.  In this
note, however, we set aside these difficulties and turn our attention
to the particle in one dimension, where the expansion is simple and
elegant.

\section{The Perturbative Expansion}

In the limit that the potential is very short range, we expect that
it can be approximated by a $\delta$-function:
\begin{equation}
V(x)\rightarrow v\,\delta(x)
\label{deltafunction}
\end{equation}
where
\begin{equation}
v=\int\,dx\,V(x)\,.
\label{deltafunction2}
\end{equation}
The $\delta$-function potential has a single bound-state, of the form
\begin{equation}
\psi(x)\propto e^{-\kappa|x|}
\label{deltafunction3}
\end{equation}
where
\begin{equation}
\kappa=-mv
\label{deltafunction4}
\end{equation}
We seek a systematic expansion in which
(\ref{deltafunction})-(\ref{deltafunction4}) is the first term. Note that
because $\kappa\propto V(x)$, we expect that an expansion in powers of $\kappa
a$ will be an expansion in powers of $V$ as well. A simple
expansion of the Fourier transform of $V(x)$ in powers of the momentum
is not suitable because the higher order terms, derivatives of
$\delta$-functions, are too singular. Instead, we will exploit the relation
between the ground state of the $\delta$-function potential and the free
particle Green's function, $G_{\kappa}$:
\begin{equation}
G_{\kappa}(x) = {e^{-\kappa|x|}\over 2\kappa}.
\label{greens}
\end{equation}

In integral form, the Schr\" odinger equation reads
\begin{equation}
\psi(x) = - \int dx'\, G_{\kappa}(x-x')\, 2mV(x')\psi(x')
\label{schrodinger}
\end{equation}
where $\kappa$ is related to the energy of the state, $E_0$, by
\begin{equation}
\kappa=\sqrt{-2mE_0}\,.
\label{kappadef}
\end{equation}
The weakly bound regime occurs when $\kappa a\ll 1$ where $a$ is the range of
the potential.

If the argument of the Green's function is small compared to $1/\kappa$, we
can legitimately expand the
exponential in equation \ref{greens}. In general, in (\ref{schrodinger}), this
is not true. However, if we multiply (\ref{schrodinger}) by $V(x)$, then
we get an integral equation for the product,
$V(x)\psi(x)$:
\begin{equation}
V(x)\psi(x) = -V(x)\int dx'\, G_{\kappa}(x-x')\, 2mV(x')\psi(x')\,.
\label{vpsi}
\end{equation}
In (\ref{vpsi}), the short range of the potential guarantees that can
do this expansion because the $x-x'$ in the Green's function is sandwiched
between $V(x)$ and $V(x')$, so $|x-x'|$ cannot get much larger than $a$.

By adding and subtracting 1 from the exponential in $G_\kappa$, we can rewrite
(\ref{vpsi}) as
\begin{equation}
\begin{array}{rcl}
V(x)\psi(x) & = & -{mV(x)\over\kappa} \int dx'\, V(x')\psi(x') \\
& & - {mV(x)\over\kappa}
\int dx'\, \left[ e^{-\kappa|x-x'|}-1\right]\,
V(x')\psi(x').
\end{array}
\label{vpsi2}
\end{equation}
Since the principal contribution to the second integral comes only
from the region $x,x'\le a$, the second term on the right hand side of
(\ref{vpsi2}) is explicitly suppressed
by a factor of $\kappa a$ relative to the first.  The expression $\int
dx'\, V(x') \psi(x')$ appears as a factor in all the terms when
$V(x)\psi(x)$ is expanded to an arbitrary order so it is to convenient
to work from the start instead with the normalized function
\begin{equation}
\chi(x) \equiv {V(x)\psi(x)\over \int dx'\, V(x')\psi(x') }
\label{chi}
\end{equation}
satisfying
\begin{equation}
\int dx\, \chi(x)=1\,.
\label{chinormalization}
\end{equation}
In terms of $\chi$, we have
\begin{equation}
\chi(x) = - {mV(x)\over\kappa} - {mV(x)\over\kappa}\int dx' \left[
e^{-\kappa|x-x'|}-1\right]\, \chi(x') .
\label{iterate}
\end{equation}
This expression for $\chi(x)$ serves as the basis for our perturbation
series.

\subsection{The Energy}

{}From the power series for $\chi(x)$ in equation \ref{iterate}, we can
obtain an expression for the bound state energy by applying the
normalization condition, (\ref{chinormalization}).  For example, simply
neglecting the second term in (\ref{iterate}) and imposing
(\ref{chinormalization}) yields
\begin{equation}
1=\int dx\,{mV(x)\over\kappa}
\label{firstorder}
\end{equation}
which is equivalent to (\ref{deltafunction4}). Indeed, for a $\delta$-function
potential, the second term in (\ref{iterate}) vanishes, thus
(\ref{firstorder}) is exact, and we reproduce the solution
(\ref{deltafunction3}).

Iterating (\ref{iterate}) once and neglecting terms of order $m^3V^3$ yields
\begin{equation}
\chi(x)  =  - {mV(x)\over\kappa} + 2mV(x)\int dx_1 \left[
{e^{-\kappa|x-x_1|}-1\over 2\kappa} \right]\, {mV(x_1)\over\kappa}  \,.
\end{equation}
The leading contribution to $\chi$ is
\begin{equation}
\chi(x) = - {mV(x)\over\kappa} - {m^2V(x)\over \kappa} \int dx_1\, |x-x_1|\,
V(x_1) + \cdots
\end{equation}
for which the normalization condition determines the energy to be
\begin{equation}
\kappa = -m\int dx\, V(x) - m^2 \int\, dx_1 dx_2 |x_1-x_2|\, V(x_1)V(x_2)
+ \cdots .
\label{kappa2}
\end{equation}
It is straightforward to continue the expansion to higher order.
The expression up to fourth order for $\kappa$ is given in the appendix, in
equation \ref{kappa}.

Although the terms begin to proliferate as we expand to higher orders,
the set of integrals that contribute with an overall coefficient of
$m^n$ can be described.  The set consists of all {\it inequivalent\/}
integrals that contain $n$ integrations, $n$ factors of the potential,
and $n-1$ factors of the form $|x_j-x_{j+1}|^{m_j}$ where the sum of
the exponents satisfies $\sum_{j=1}^{n-1} m_j = n-1$.  Since the
$|x_j-x_{j+1}|^{m_j}$ arises from our expansion of the Green's
function, (\ref{greens}), it is accompanied by a $(1/m_j!)$.  Thus,
general $n$-th order term looks like
\begin{equation}
-m^ns \int \left( \prod_{i=1}^n dx_i\, V(x_i) \right)
\left( \prod_{j=1}^{n-1} {1\over m_j!} |x_i-x_{i+1}|^{m_j} \right) .
\label{nthorder}
\end{equation}
There only remains the factor $s$ to explain.  In demanding that the
integrals be inequivalent, we have slightly undercounted, and the $s$
corrects for this omission.  A specific contribution to
(\ref{nthorder}) often contains several connected integrals and with
each such piece written as
\begin{equation}
\int dx_1\cdots dx_{k\le n}\,
|x_1-x_2|^{m_1} |x_2-x_3|^{m_2}\cdots |x_{k-1}-x_k|^{m_{k-1}}
V(x_1)\cdots V(x_k) .
\label{connected}
\end{equation}
$s-1$, then, counts the number of non-trivial ways of rearranging the
$m_j$'s within this connected factor which yield different integrands
but the same integral.  Similarly, if our term (\ref{nthorder})
contains several connected integrals, the total $s$ is the product of
these symmetry factors for each connected component.  For example, at
fourth order, shown in appendix equation (\ref{kappa}), we have a
term of the form,
\begin{equation}
\Biggl(\int dx\, V(x)\Biggr) \Biggl(
\int dx_1\,dx_2\,dx_3\, |x_1-x_2|^2 |x_2-x_3|\, V(x_1)V(x_2)V(x_3) \Biggr) .
\label{symexample}
\end{equation}
The exponents of the second connected piece could be switched so that the
integrand has $|x_1-x_2| |x_2-x_3|^2$ instead of $|x_1-x_2|^2 |x_2-x_3|$;
but both integrals are equivalent.  Thus for (\ref{symexample}), $s=2$ (which
cancels the $1/2!$ from the Green's function expansion).

\subsection{The Wave Function}

The expression for $\chi(x)$ to any order contains an explicit factor of
$V(x)$. One might think, from (\ref{chi}), that we could simply cancel this
factor and get the wave function up to a normalization factor. This is not
quite right, because the resulting series only converges for small $x$.
However, we can construct the wave function for all $x$ using
(\ref{schrodinger}), which can be written as
\begin{equation}
\psi(x)=N\int dx'\,e^{-\kappa|x-x'|}\,\chi(x')
\label{psi}
\end{equation}
where $N$ is a normalization factor.

\section{Example: The Square Well Potential}

For the square well potential,
\begin{equation}
V(x) = \left\{ \begin{array}{ll}
-V_0 & \mbox{for $|x|\le a$}\\
0 &\mbox{for $|x|>a$}
\end{array}
\right.
\label{square}
\end{equation}
our expansion is in powers of the dimensionless quantity,
\begin{equation}
2mV_0a^2\,.
\label{square2}
\end{equation}
If we apply (\ref{kappa}), we find
\begin{equation}
\begin{array}{c}
\kappa a  =  (2mV_0a^2) - {2\over 3}(2mV_0a^2)^2 + {4\over 5} (2mV_0a^2)^3
\\
 - {368\over 315} (2mV_0a^2)^4 + {\cal O}\left( (2mV_0a^2)^5 \right)
+\cdots\, .
\end{array}
\label{square3}
\end{equation}
which agrees with what we get from a Taylor expansion of the exact value of
$\kappa$.

Rather than show the rather complicated formulae for the expansion
of the wave function, we will display, in figures 1-4, plots of the exact wave
function along with the first four approximates, for various values of
$2mV_0a^2$. Evidently, for $2mV_0a^2=0.3$, shown in figure 1, the expansion
for the wave function converges rapidly, while for $2mV_0a^2=0.9$, shown in
figure 4, the expansion is not working at all.

{\figsize\begin{figure}[f]
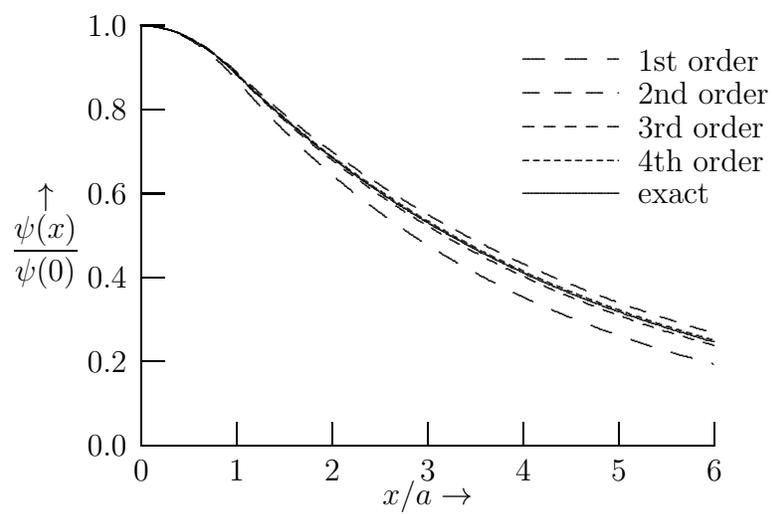

$$\beginpicture
\setcoordinatesystem units <.5truein,2.2truein>
\setplotarea x from 0 to 6, y from 0 to 1
\axis bottom visible ticks in numbered from 0 to 6 by 1 /
\axis left visible ticks in numbered from 0.0 to 1.0 by .2 /
\put {$\displaystyle \uparrow\atop\displaystyle
{\psi(x)\over\psi(0)}$} at -1 .5
\put {$x/a \rightarrow$} at 3 -.12
\setdashes <8pt>
\plot 4 .92 5 .92 /
\put {1st order} [l] at 5.2 .92
\setdashes <6pt>
\plot 4 .84 5 .84 /
\put {2nd order} [l] at 5.2 .84
\setdashes <4pt>
\plot 4 .76 5 .76 /
\put {3rd order} [l] at 5.2 .76
\setdashes <2pt>
\plot 4 .68 5 .68 /
\put {4th order} [l] at 5.2 .68
\setsolid
\plot 4 .6 5 .6 /
\put {exact} [l] at 5.2 .6
\setdashes <8pt>
\plot
   0 1.000 .04 1.000 .08 .999 .12 .998 .16 .997
       .20 .995 .24 .993 .28 .990 .32 .987 .36 .983
       .40 .979 .44 .975 .48 .970 .52 .965 .56 .960
       .60 .954 .64 .947 .68 .940 .72 .933 .76 .925
       .80 .917 .84 .909 .88 .900 .92 .890 .96 .881
       1.00 .870 1.04 .860 1.08 .850 1.12 .840 1.16 .830
       1.20 .820 1.24 .810 1.28 .800 1.32 .791 1.36 .781
       1.40 .772 1.44 .763 1.48 .754 1.52 .745 1.56 .736
       1.60 .727 1.64 .718 1.68 .710 1.72 .701 1.76 .693
       1.80 .685 1.84 .677 1.88 .668 1.92 .660 1.96 .653
       2.00 .645 2.04 .637 2.08 .630 2.12 .622 2.16 .615
       2.20 .607 2.24 .600 2.28 .593 2.32 .586 2.36 .579
       2.40 .572 2.44 .565 2.48 .558 2.52 .552 2.56 .545
       2.60 .539 2.64 .532 2.68 .526 2.72 .520 2.76 .513
       2.80 .507 2.84 .501 2.88 .495 2.92 .489 2.96 .483
       3.00 .478 3.04 .472 3.08 .466 3.12 .461 3.16 .455
       3.20 .450 3.24 .445 3.28 .439 3.32 .434 3.36 .429
       3.40 .424 3.44 .419 3.48 .414 3.52 .409 3.56 .404
       3.60 .399 3.64 .394 3.68 .390 3.72 .385 3.76 .380
       3.80 .376 3.84 .371 3.88 .367 3.92 .362 3.96 .358
       4.00 .354 4.04 .350 4.08 .345 4.12 .341 4.16 .337
       4.20 .333 4.24 .329 4.28 .325 4.32 .321 4.36 .318
       4.40 .314 4.44 .310 4.48 .306 4.52 .303 4.56 .299
       4.60 .296 4.64 .292 4.68 .289 4.72 .285 4.76 .282
       4.80 .278 4.84 .275 4.88 .272 4.92 .269 4.96 .265
       5.00 .262 5.04 .259 5.08 .256 5.12 .253 5.16 .250
       5.20 .247 5.24 .244 5.28 .241 5.32 .238 5.36 .235
       5.40 .233 5.44 .230 5.48 .227 5.52 .224 5.56 .222
       5.60 .219 5.64 .216 5.68 .214 5.72 .211 5.76 .209
       5.80 .206 5.84 .204 5.88 .201 5.92 .199 5.96 .197
       6.00 .194 /
\setdashes <6pt>
\plot
0 1.000 .04 1.000 .08 .999 .12 .998 .16 .997
       .20 .995 .24 .993 .28 .991 .32 .988 .36 .985
       .40 .982 .44 .978 .48 .974 .52 .969 .56 .965
       .60 .959 .64 .954 .68 .948 .72 .942 .76 .935
       .80 .928 .84 .921 .88 .914 .92 .906 .96 .898
       1.00 .890 1.04 .881 1.08 .873 1.12 .864 1.16 .856
       1.20 .848 1.24 .840 1.28 .832 1.32 .824 1.36 .816
       1.40 .808 1.44 .800 1.48 .793 1.52 .785 1.56 .778
       1.60 .770 1.64 .763 1.68 .756 1.72 .748 1.76 .741
       1.80 .734 1.84 .727 1.88 .720 1.92 .713 1.96 .707
       2.00 .700 2.04 .693 2.08 .686 2.12 .680 2.16 .673
       2.20 .667 2.24 .661 2.28 .654 2.32 .648 2.36 .642
       2.40 .636 2.44 .630 2.48 .624 2.52 .618 2.56 .612
       2.60 .606 2.64 .600 2.68 .594 2.72 .589 2.76 .583
       2.80 .578 2.84 .572 2.88 .567 2.92 .561 2.96 .556
       3.00 .550 3.04 .545 3.08 .540 3.12 .535 3.16 .530
       3.20 .525 3.24 .520 3.28 .515 3.32 .510 3.36 .505
       3.40 .500 3.44 .495 3.48 .491 3.52 .486 3.56 .481
       3.60 .477 3.64 .472 3.68 .468 3.72 .463 3.76 .459
       3.80 .454 3.84 .450 3.88 .446 3.92 .441 3.96 .437
       4.00 .433 4.04 .429 4.08 .425 4.12 .421 4.16 .417
       4.20 .413 4.24 .409 4.28 .405 4.32 .401 4.36 .397
       4.40 .393 4.44 .390 4.48 .386 4.52 .382 4.56 .379
       4.60 .375 4.64 .371 4.68 .368 4.72 .364 4.76 .361
       4.80 .357 4.84 .354 4.88 .351 4.92 .347 4.96 .344
       5.00 .341 5.04 .337 5.08 .334 5.12 .331 5.16 .328
       5.20 .325 5.24 .322 5.28 .318 5.32 .315 5.36 .312
       5.40 .309 5.44 .306 5.48 .304 5.52 .301 5.56 .298
       5.60 .295 5.64 .292 5.68 .289 5.72 .287 5.76 .284
       5.80 .281 5.84 .278 5.88 .276 5.92 .273 5.96 .271
       6.00 .268 /
\setdashes <4pt>
\plot
   0 1.000 .04 1.000 .08 .999 .12 .998 .16 .997
       .20 .995 .24 .993 .28 .991 .32 .988 .36 .985
       .40 .981 .44 .977 .48 .973 .52 .968 .56 .963
       .60 .957 .64 .952 .68 .945 .72 .939 .76 .932
       .80 .925 .84 .917 .88 .909 .92 .901 .96 .892
       1.00 .883 1.04 .874 1.08 .865 1.12 .856 1.16 .847
       1.20 .838 1.24 .829 1.28 .821 1.32 .812 1.36 .804
       1.40 .795 1.44 .787 1.48 .779 1.52 .771 1.56 .763
       1.60 .755 1.64 .747 1.68 .739 1.72 .731 1.76 .724
       1.80 .716 1.84 .709 1.88 .701 1.92 .694 1.96 .687
       2.00 .680 2.04 .673 2.08 .666 2.12 .659 2.16 .652
       2.20 .645 2.24 .638 2.28 .632 2.32 .625 2.36 .619
       2.40 .612 2.44 .606 2.48 .599 2.52 .593 2.56 .587
       2.60 .581 2.64 .575 2.68 .569 2.72 .563 2.76 .557
       2.80 .551 2.84 .546 2.88 .540 2.92 .534 2.96 .529
       3.00 .523 3.04 .518 3.08 .512 3.12 .507 3.16 .502
       3.20 .497 3.24 .491 3.28 .486 3.32 .481 3.36 .476
       3.40 .471 3.44 .466 3.48 .461 3.52 .457 3.56 .452
       3.60 .447 3.64 .443 3.68 .438 3.72 .433 3.76 .429
       3.80 .424 3.84 .420 3.88 .416 3.92 .411 3.96 .407
       4.00 .403 4.04 .399 4.08 .394 4.12 .390 4.16 .386
       4.20 .382 4.24 .378 4.28 .374 4.32 .370 4.36 .367
       4.40 .363 4.44 .359 4.48 .355 4.52 .352 4.56 .348
       4.60 .344 4.64 .341 4.68 .337 4.72 .334 4.76 .330
       4.80 .327 4.84 .323 4.88 .320 4.92 .317 4.96 .313
       5.00 .310 5.04 .307 5.08 .304 5.12 .301 5.16 .297
       5.20 .294 5.24 .291 5.28 .288 5.32 .285 5.36 .282
       5.40 .279 5.44 .276 5.48 .273 5.52 .271 5.56 .268
       5.60 .265 5.64 .262 5.68 .260 5.72 .257 5.76 .254
       5.80 .252 5.84 .249 5.88 .246 5.92 .244 5.96 .241
       6.00 .239 /
\setdashes <2pt>
\plot
   0 1.000 .04 1.000 .08 .999 .12 .998 .16 .997
       .20 .995 .24 .993 .28 .991 .32 .988 .36 .985
       .40 .981 .44 .977 .48 .973 .52 .969 .56 .964
       .60 .958 .64 .953 .68 .947 .72 .940 .76 .933
       .80 .926 .84 .919 .88 .911 .92 .903 .96 .895
       1.00 .886 1.04 .877 1.08 .868 1.12 .859 1.16 .851
       1.20 .842 1.24 .834 1.28 .825 1.32 .817 1.36 .809
       1.40 .801 1.44 .793 1.48 .785 1.52 .777 1.56 .769
       1.60 .761 1.64 .754 1.68 .746 1.72 .739 1.76 .731
       1.80 .724 1.84 .717 1.88 .709 1.92 .702 1.96 .695
       2.00 .688 2.04 .681 2.08 .675 2.12 .668 2.16 .661
       2.20 .654 2.24 .648 2.28 .641 2.32 .635 2.36 .629
       2.40 .622 2.44 .616 2.48 .610 2.52 .604 2.56 .598
       2.60 .592 2.64 .586 2.68 .580 2.72 .574 2.76 .568
       2.80 .563 2.84 .557 2.88 .551 2.92 .546 2.96 .540
       3.00 .535 3.04 .530 3.08 .524 3.12 .519 3.16 .514
       3.20 .509 3.24 .504 3.28 .498 3.32 .493 3.36 .489
       3.40 .484 3.44 .479 3.48 .474 3.52 .469 3.56 .464
       3.60 .460 3.64 .455 3.68 .451 3.72 .446 3.76 .442
       3.80 .437 3.84 .433 3.88 .428 3.92 .424 3.96 .420
       4.00 .416 4.04 .412 4.08 .407 4.12 .403 4.16 .399
       4.20 .395 4.24 .391 4.28 .387 4.32 .383 4.36 .380
       4.40 .376 4.44 .372 4.48 .368 4.52 .365 4.56 .361
       4.60 .357 4.64 .354 4.68 .350 4.72 .347 4.76 .343
       4.80 .340 4.84 .336 4.88 .333 4.92 .330 4.96 .326
       5.00 .323 5.04 .320 5.08 .317 5.12 .313 5.16 .310
       5.20 .307 5.24 .304 5.28 .301 5.32 .298 5.36 .295
       5.40 .292 5.44 .289 5.48 .286 5.52 .283 5.56 .281
       5.60 .278 5.64 .275 5.68 .272 5.72 .269 5.76 .267
       5.80 .264 5.84 .261 5.88 .259 5.92 .256 5.96 .254
       6.00 .251 /
\setsolid
\plot
   0 1.000 .04 1.000 .08 .999 .12 .998 .16 .997
       .20 .995 .24 .993 .28 .991 .32 .988 .36 .985
       .40 .981 .44 .977 .48 .973 .52 .968 .56 .963
       .60 .958 .64 .952 .68 .946 .72 .940 .76 .933
       .80 .926 .84 .918 .88 .910 .92 .902 .96 .894
       1.00 .885 1.04 .876 1.08 .867 1.12 .858 1.16 .849
       1.20 .841 1.24 .832 1.28 .824 1.32 .815 1.36 .807
       1.40 .799 1.44 .791 1.48 .783 1.52 .775 1.56 .767
       1.60 .759 1.64 .752 1.68 .744 1.72 .736 1.76 .729
       1.80 .721 1.84 .714 1.88 .707 1.92 .700 1.96 .693
       2.00 .686 2.04 .679 2.08 .672 2.12 .665 2.16 .658
       2.20 .651 2.24 .645 2.28 .638 2.32 .632 2.36 .625
       2.40 .619 2.44 .613 2.48 .607 2.52 .600 2.56 .594
       2.60 .588 2.64 .582 2.68 .576 2.72 .570 2.76 .565
       2.80 .559 2.84 .553 2.88 .548 2.92 .542 2.96 .537
       3.00 .531 3.04 .526 3.08 .520 3.12 .515 3.16 .510
       3.20 .505 3.24 .500 3.28 .495 3.32 .490 3.36 .485
       3.40 .480 3.44 .475 3.48 .470 3.52 .465 3.56 .460
       3.60 .456 3.64 .451 3.68 .447 3.72 .442 3.76 .438
       3.80 .433 3.84 .429 3.88 .424 3.92 .420 3.96 .416
       4.00 .412 4.04 .407 4.08 .403 4.12 .399 4.16 .395
       4.20 .391 4.24 .387 4.28 .383 4.32 .379 4.36 .375
       4.40 .372 4.44 .368 4.48 .364 4.52 .360 4.56 .357
       4.60 .353 4.64 .350 4.68 .346 4.72 .342 4.76 .339
       4.80 .336 4.84 .332 4.88 .329 4.92 .325 4.96 .322
       5.00 .319 5.04 .316 5.08 .312 5.12 .309 5.16 .306
       5.20 .303 5.24 .300 5.28 .297 5.32 .294 5.36 .291
       5.40 .288 5.44 .285 5.48 .282 5.52 .279 5.56 .276
       5.60 .274 5.64 .271 5.68 .268 5.72 .265 5.76 .263
       5.80 .260 5.84 .257 5.88 .255 5.92 .252 5.96 .250
       6.00 .247 /
\endpicture$$
\caption{\label{fig-1} $2mV_0a^2=0.3$
}\end{figure}}

{\figsize\begin{figure}[f]
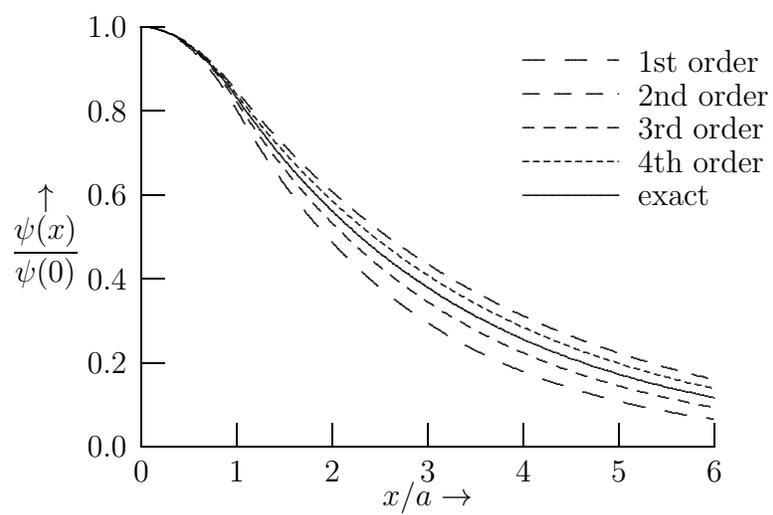

$$\beginpicture
\setcoordinatesystem units <.5truein,2.2truein>
\setplotarea x from 0 to 6, y from 0 to 1
\axis bottom visible ticks in numbered from 0 to 6 by 1 /
\axis left visible ticks in numbered from 0.0 to 1.0 by .2 /
\put {$\displaystyle \uparrow\atop\displaystyle
{\psi(x)\over\psi(0)}$} at -1 .5
\put {$x/a \rightarrow$} at 3 -.12
\setdashes <8pt>
\plot 4 .92 5 .92 /
\put {1st order} [l] at 5.2 .92
\setdashes <6pt>
\plot 4 .84 5 .84 /
\put {2nd order} [l] at 5.2 .84
\setdashes <4pt>
\plot 4 .76 5 .76 /
\put {3rd order} [l] at 5.2 .76
\setdashes <2pt>
\plot 4 .68 5 .68 /
\put {4th order} [l] at 5.2 .68
\setsolid
\plot 4 .6 5 .6 /
\put {exact} [l] at 5.2 .6
\setdashes <8pt>
\plot
  0 1.000 .04 1.000 .08 .999 .12 .997 .16 .995
       .20 .992 .24 .989 .28 .985 .32 .980 .36 .975
       .40 .969 .44 .963 .48 .955 .52 .948 .56 .939
       .60 .930 .64 .920 .68 .910 .72 .899 .76 .887
       .80 .875 .84 .862 .88 .848 .92 .834 .96 .819
       1.00 .803 1.04 .787 1.08 .772 1.12 .756 1.16 .742
       1.20 .727 1.24 .712 1.28 .698 1.32 .684 1.36 .671
       1.40 .658 1.44 .645 1.48 .632 1.52 .619 1.56 .607
       1.60 .595 1.64 .583 1.68 .572 1.72 .560 1.76 .549
       1.80 .538 1.84 .528 1.88 .517 1.92 .507 1.96 .497
       2.00 .487 2.04 .478 2.08 .468 2.12 .459 2.16 .450
       2.20 .441 2.24 .432 2.28 .424 2.32 .415 2.36 .407
       2.40 .399 2.44 .391 2.48 .383 2.52 .376 2.56 .368
       2.60 .361 2.64 .354 2.68 .347 2.72 .340 2.76 .333
       2.80 .327 2.84 .320 2.88 .314 2.92 .308 2.96 .301
       3.00 .296 3.04 .290 3.08 .284 3.12 .278 3.16 .273
       3.20 .267 3.24 .262 3.28 .257 3.32 .252 3.36 .247
       3.40 .242 3.44 .237 3.48 .232 3.52 .228 3.56 .223
       3.60 .219 3.64 .215 3.68 .210 3.72 .206 3.76 .202
       3.80 .198 3.84 .194 3.88 .190 3.92 .187 3.96 .183
       4.00 .179 4.04 .176 4.08 .172 4.12 .169 4.16 .165
       4.20 .162 4.24 .159 4.28 .156 4.32 .153 4.36 .150
       4.40 .147 4.44 .144 4.48 .141 4.52 .138 4.56 .135
       4.60 .133 4.64 .130 4.68 .128 4.72 .125 4.76 .123
       4.80 .120 4.84 .118 4.88 .115 4.92 .113 4.96 .111
       5.00 .109 5.04 .107 5.08 .104 5.12 .102 5.16 .100
       5.20 .098 5.24 .096 5.28 .095 5.32 .093 5.36 .091
       5.40 .089 5.44 .087 5.48 .086 5.52 .084 5.56 .082
       5.60 .081 5.64 .079 5.68 .077 5.72 .076 5.76 .074
       5.80 .073 5.84 .071 5.88 .070 5.92 .069 5.96 .067
       6.00 .066 /
\setdashes <6pt>
\plot
   0 1.000 .04 1.000 .08 .999 .12 .998 .16 .996
       .20 .994 .24 .991 .28 .987 .32 .983 .36 .979
       .40 .974 .44 .969 .48 .963 .52 .957 .56 .950
       .60 .943 .64 .935 .68 .927 .72 .919 .76 .910
       .80 .900 .84 .891 .88 .880 .92 .870 .96 .859
       1.00 .848 1.04 .837 1.08 .826 1.12 .815 1.16 .804
       1.20 .793 1.24 .783 1.28 .772 1.32 .762 1.36 .752
       1.40 .742 1.44 .732 1.48 .723 1.52 .713 1.56 .704
       1.60 .694 1.64 .685 1.68 .676 1.72 .667 1.76 .658
       1.80 .649 1.84 .641 1.88 .632 1.92 .624 1.96 .616
       2.00 .608 2.04 .600 2.08 .592 2.12 .584 2.16 .576
       2.20 .568 2.24 .561 2.28 .553 2.32 .546 2.36 .539
       2.40 .532 2.44 .525 2.48 .518 2.52 .511 2.56 .504
       2.60 .497 2.64 .491 2.68 .484 2.72 .478 2.76 .472
       2.80 .465 2.84 .459 2.88 .453 2.92 .447 2.96 .441
       3.00 .435 3.04 .430 3.08 .424 3.12 .418 3.16 .413
       3.20 .407 3.24 .402 3.28 .397 3.32 .391 3.36 .386
       3.40 .381 3.44 .376 3.48 .371 3.52 .366 3.56 .361
       3.60 .356 3.64 .352 3.68 .347 3.72 .342 3.76 .338
       3.80 .333 3.84 .329 3.88 .325 3.92 .320 3.96 .316
       4.00 .312 4.04 .308 4.08 .304 4.12 .300 4.16 .296
       4.20 .292 4.24 .288 4.28 .284 4.32 .280 4.36 .277
       4.40 .273 4.44 .269 4.48 .266 4.52 .262 4.56 .259
       4.60 .255 4.64 .252 4.68 .249 4.72 .245 4.76 .242
       4.80 .239 4.84 .236 4.88 .233 4.92 .230 4.96 .227
       5.00 .224 5.04 .221 5.08 .218 5.12 .215 5.16 .212
       5.20 .209 5.24 .206 5.28 .204 5.32 .201 5.36 .198
       5.40 .196 5.44 .193 5.48 .190 5.52 .188 5.56 .185
       5.60 .183 5.64 .181 5.68 .178 5.72 .176 5.76 .174
       5.80 .171 5.84 .169 5.88 .167 5.92 .164 5.96 .162
       6.00 .160 /
\setdashes <4pt>
\plot
   0 1.000 .04 1.000 .08 .999 .12 .997 .16 .995
       .20 .993 .24 .990 .28 .986 .32 .982 .36 .977
       .40 .971 .44 .966 .48 .959 .52 .952 .56 .944
       .60 .936 .64 .927 .68 .918 .72 .908 .76 .897
       .80 .886 .84 .874 .88 .862 .92 .849 .96 .836
       1.00 .822 1.04 .808 1.08 .794 1.12 .780 1.16 .767
       1.20 .754 1.24 .741 1.28 .728 1.32 .716 1.36 .703
       1.40 .691 1.44 .679 1.48 .668 1.52 .656 1.56 .645
       1.60 .634 1.64 .623 1.68 .612 1.72 .602 1.76 .591
       1.80 .581 1.84 .571 1.88 .561 1.92 .552 1.96 .542
       2.00 .533 2.04 .524 2.08 .515 2.12 .506 2.16 .497
       2.20 .489 2.24 .480 2.28 .472 2.32 .464 2.36 .456
       2.40 .448 2.44 .440 2.48 .433 2.52 .425 2.56 .418
       2.60 .411 2.64 .404 2.68 .397 2.72 .390 2.76 .383
       2.80 .377 2.84 .370 2.88 .364 2.92 .358 2.96 .352
       3.00 .345 3.04 .340 3.08 .334 3.12 .328 3.16 .322
       3.20 .317 3.24 .311 3.28 .306 3.32 .301 3.36 .296
       3.40 .291 3.44 .286 3.48 .281 3.52 .276 3.56 .271
       3.60 .266 3.64 .262 3.68 .257 3.72 .253 3.76 .249
       3.80 .244 3.84 .240 3.88 .236 3.92 .232 3.96 .228
       4.00 .224 4.04 .220 4.08 .216 4.12 .213 4.16 .209
       4.20 .205 4.24 .202 4.28 .198 4.32 .195 4.36 .192
       4.40 .188 4.44 .185 4.48 .182 4.52 .179 4.56 .176
       4.60 .173 4.64 .170 4.68 .167 4.72 .164 4.76 .161
       4.80 .158 4.84 .156 4.88 .153 4.92 .150 4.96 .148
       5.00 .145 5.04 .143 5.08 .140 5.12 .138 5.16 .136
       5.20 .133 5.24 .131 5.28 .129 5.32 .126 5.36 .124
       5.40 .122 5.44 .120 5.48 .118 5.52 .116 5.56 .114
       5.60 .112 5.64 .110 5.68 .108 5.72 .106 5.76 .104
       5.80 .103 5.84 .101 5.88 .099 5.92 .097 5.96 .096
       6.00 .094 /
\setdashes <2pt>
\plot
   0 1.000 .04 1.000 .08 .999 .12 .998 .16 .996
       .20 .993 .24 .990 .28 .987 .32 .983 .36 .978
       .40 .973 .44 .968 .48 .962 .52 .955 .56 .948
       .60 .940 .64 .932 .68 .924 .72 .915 .76 .905
       .80 .896 .84 .885 .88 .875 .92 .863 .96 .852
       1.00 .840 1.04 .828 1.08 .816 1.12 .804 1.16 .793
       1.20 .782 1.24 .770 1.28 .759 1.32 .748 1.36 .738
       1.40 .727 1.44 .717 1.48 .707 1.52 .696 1.56 .686
       1.60 .677 1.64 .667 1.68 .657 1.72 .648 1.76 .639
       1.80 .630 1.84 .621 1.88 .612 1.92 .603 1.96 .594
       2.00 .586 2.04 .577 2.08 .569 2.12 .561 2.16 .553
       2.20 .545 2.24 .537 2.28 .530 2.32 .522 2.36 .515
       2.40 .507 2.44 .500 2.48 .493 2.52 .486 2.56 .479
       2.60 .472 2.64 .465 2.68 .459 2.72 .452 2.76 .445
       2.80 .439 2.84 .433 2.88 .427 2.92 .421 2.96 .415
       3.00 .409 3.04 .403 3.08 .397 3.12 .391 3.16 .386
       3.20 .380 3.24 .375 3.28 .369 3.32 .364 3.36 .359
       3.40 .354 3.44 .349 3.48 .344 3.52 .339 3.56 .334
       3.60 .329 3.64 .324 3.68 .320 3.72 .315 3.76 .311
       3.80 .306 3.84 .302 3.88 .298 3.92 .293 3.96 .289
       4.00 .285 4.04 .281 4.08 .277 4.12 .273 4.16 .269
       4.20 .265 4.24 .261 4.28 .258 4.32 .254 4.36 .250
       4.40 .247 4.44 .243 4.48 .240 4.52 .236 4.56 .233
       4.60 .230 4.64 .226 4.68 .223 4.72 .220 4.76 .217
       4.80 .214 4.84 .211 4.88 .208 4.92 .205 4.96 .202
       5.00 .199 5.04 .196 5.08 .193 5.12 .190 5.16 .188
       5.20 .185 5.24 .182 5.28 .180 5.32 .177 5.36 .175
       5.40 .172 5.44 .170 5.48 .167 5.52 .165 5.56 .162
       5.60 .160 5.64 .158 5.68 .156 5.72 .153 5.76 .151
       5.80 .149 5.84 .147 5.88 .145 5.92 .143 5.96 .141
       6.00 .139 /
\setsolid
\plot
   0 1.000 .04 1.000 .08 .999 .12 .998 .16 .996
       .20 .993 .24 .990 .28 .986 .32 .982 .36 .978
       .40 .972 .44 .967 .48 .960 .52 .954 .56 .946
       .60 .938 .64 .930 .68 .921 .72 .912 .76 .902
       .80 .891 .84 .880 .88 .869 .92 .857 .96 .845
       1.00 .832 1.04 .819 1.08 .806 1.12 .794 1.16 .781
       1.20 .769 1.24 .757 1.28 .745 1.32 .734 1.36 .722
       1.40 .711 1.44 .700 1.48 .689 1.52 .678 1.56 .668
       1.60 .657 1.64 .647 1.68 .637 1.72 .627 1.76 .617
       1.80 .608 1.84 .598 1.88 .589 1.92 .580 1.96 .571
       2.00 .562 2.04 .553 2.08 .545 2.12 .536 2.16 .528
       2.20 .520 2.24 .511 2.28 .503 2.32 .496 2.36 .488
       2.40 .480 2.44 .473 2.48 .465 2.52 .458 2.56 .451
       2.60 .444 2.64 .437 2.68 .430 2.72 .424 2.76 .417
       2.80 .411 2.84 .404 2.88 .398 2.92 .392 2.96 .386
       3.00 .380 3.04 .374 3.08 .368 3.12 .362 3.16 .356
       3.20 .351 3.24 .345 3.28 .340 3.32 .335 3.36 .330
       3.40 .324 3.44 .319 3.48 .314 3.52 .309 3.56 .305
       3.60 .300 3.64 .295 3.68 .291 3.72 .286 3.76 .282
       3.80 .277 3.84 .273 3.88 .269 3.92 .265 3.96 .260
       4.00 .256 4.04 .252 4.08 .248 4.12 .245 4.16 .241
       4.20 .237 4.24 .233 4.28 .230 4.32 .226 4.36 .223
       4.40 .219 4.44 .216 4.48 .212 4.52 .209 4.56 .206
       4.60 .203 4.64 .199 4.68 .196 4.72 .193 4.76 .190
       4.80 .187 4.84 .184 4.88 .182 4.92 .179 4.96 .176
       5.00 .173 5.04 .170 5.08 .168 5.12 .165 5.16 .163
       5.20 .160 5.24 .158 5.28 .155 5.32 .153 5.36 .150
       5.40 .148 5.44 .146 5.48 .143 5.52 .141 5.56 .139
       5.60 .137 5.64 .135 5.68 .133 5.72 .131 5.76 .129
       5.80 .127 5.84 .125 5.88 .123 5.92 .121 5.96 .119
       6.00 .117 /
\endpicture$$
\caption{\label{fig-2} $2mV_0a^2=0.5$}
\end{figure}}

{\figsize\begin{figure}[f]
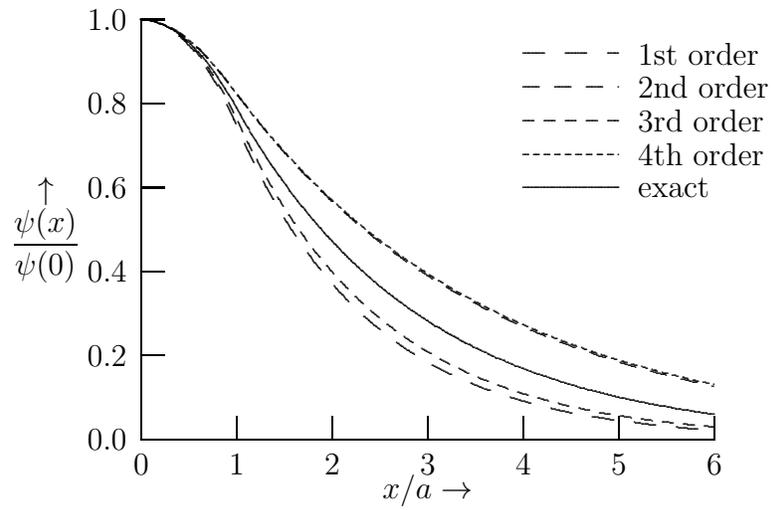

$$\beginpicture
\setcoordinatesystem units <.5truein,2.2truein>
\setplotarea x from 0 to 6, y from 0 to 1
\axis bottom visible ticks in numbered from 0 to 6 by 1 /
\axis left visible ticks in numbered from 0.0 to 1.0 by .2 /
\put {$\displaystyle \uparrow\atop\displaystyle
{\psi(x)\over\psi(0)}$} at -1 .5
\put {$x/a \rightarrow$} at 3 -.12
\setdashes <8pt>
\plot 4 .92 5 .92 /
\put {1st order} [l] at 5.2 .92
\setdashes <6pt>
\plot 4 .84 5 .84 /
\put {2nd order} [l] at 5.2 .84
\setdashes <4pt>
\plot 4 .76 5 .76 /
\put {3rd order} [l] at 5.2 .76
\setdashes <2pt>
\plot 4 .68 5 .68 /
\put {4th order} [l] at 5.2 .68
\setsolid
\plot 4 .6 5 .6 /
\put {exact} [l] at 5.2 .6
\setdashes <8pt>
\plot
0 1.000 .04 1.000 .08 .998 .12 .997 .16 .994
       .20 .990 .24 .986 .28 .981 .32 .975 .36 .969
       .40 .961 .44 .953 .48 .944 .52 .934 .56 .923
       .60 .912 .64 .899 .68 .886 .72 .872 .76 .857
       .80 .841 .84 .825 .88 .807 .92 .788 .96 .769
       1.00 .748 1.04 .728 1.08 .708 1.12 .688 1.16 .669
       1.20 .651 1.24 .633 1.28 .615 1.32 .598 1.36 .582
       1.40 .566 1.44 .550 1.48 .535 1.52 .520 1.56 .506
       1.60 .492 1.64 .478 1.68 .465 1.72 .452 1.76 .440
       1.80 .427 1.84 .416 1.88 .404 1.92 .393 1.96 .382
       2.00 .372 2.04 .361 2.08 .351 2.12 .342 2.16 .332
       2.20 .323 2.24 .314 2.28 .305 2.32 .297 2.36 .289
       2.40 .281 2.44 .273 2.48 .266 2.52 .258 2.56 .251
       2.60 .244 2.64 .237 2.68 .231 2.72 .224 2.76 .218
       2.80 .212 2.84 .206 2.88 .201 2.92 .195 2.96 .190
       3.00 .185 3.04 .179 3.08 .174 3.12 .170 3.16 .165
       3.20 .160 3.24 .156 3.28 .152 3.32 .147 3.36 .143
       3.40 .139 3.44 .136 3.48 .132 3.52 .128 3.56 .125
       3.60 .121 3.64 .118 3.68 .115 3.72 .111 3.76 .108
       3.80 .105 3.84 .102 3.88 .100 3.92 .097 3.96 .094
       4.00 .092 4.04 .089 4.08 .087 4.12 .084 4.16 .082
       4.20 .080 4.24 .077 4.28 .075 4.32 .073 4.36 .071
       4.40 .069 4.44 .067 4.48 .065 4.52 .064 4.56 .062
       4.60 .060 4.64 .059 4.68 .057 4.72 .055 4.76 .054
       4.80 .052 4.84 .051 4.88 .049 4.92 .048 4.96 .047
       5.00 .046 5.04 .044 5.08 .043 5.12 .042 5.16 .041
       5.20 .040 5.24 .038 5.28 .037 5.32 .036 5.36 .035
       5.40 .034 5.44 .033 5.48 .033 5.52 .032 5.56 .031
       5.60 .030 5.64 .029 5.68 .028 5.72 .027 5.76 .027
       5.80 .026 5.84 .025 5.88 .025 5.92 .024 5.96 .023
       6.00 .023 /
\setdashes <6pt>
\plot
   0 1.000 .04 1.000 .08 .999 .12 .997 .16 .995
       .20 .992 .24 .989 .28 .985 .32 .980 .36 .975
       .40 .969 .44 .963 .48 .956 .52 .948 .56 .940
       .60 .932 .64 .923 .68 .913 .72 .903 .76 .893
       .80 .882 .84 .871 .88 .860 .92 .848 .96 .836
       1.00 .824 1.04 .812 1.08 .800 1.12 .788 1.16 .776
       1.20 .765 1.24 .754 1.28 .742 1.32 .731 1.36 .721
       1.40 .710 1.44 .699 1.48 .689 1.52 .679 1.56 .669
       1.60 .659 1.64 .649 1.68 .639 1.72 .630 1.76 .621
       1.80 .611 1.84 .602 1.88 .593 1.92 .585 1.96 .576
       2.00 .567 2.04 .559 2.08 .551 2.12 .543 2.16 .535
       2.20 .527 2.24 .519 2.28 .511 2.32 .504 2.36 .496
       2.40 .489 2.44 .481 2.48 .474 2.52 .467 2.56 .460
       2.60 .454 2.64 .447 2.68 .440 2.72 .434 2.76 .427
       2.80 .421 2.84 .415 2.88 .409 2.92 .402 2.96 .397
       3.00 .391 3.04 .385 3.08 .379 3.12 .374 3.16 .368
       3.20 .363 3.24 .357 3.28 .352 3.32 .347 3.36 .342
       3.40 .336 3.44 .331 3.48 .327 3.52 .322 3.56 .317
       3.60 .312 3.64 .308 3.68 .303 3.72 .299 3.76 .294
       3.80 .290 3.84 .285 3.88 .281 3.92 .277 3.96 .273
       4.00 .269 4.04 .265 4.08 .261 4.12 .257 4.16 .253
       4.20 .250 4.24 .246 4.28 .242 4.32 .239 4.36 .235
       4.40 .232 4.44 .228 4.48 .225 4.52 .221 4.56 .218
       4.60 .215 4.64 .212 4.68 .209 4.72 .206 4.76 .203
       4.80 .199 4.84 .197 4.88 .194 4.92 .191 4.96 .188
       5.00 .185 5.04 .182 5.08 .180 5.12 .177 5.16 .174
       5.20 .172 5.24 .169 5.28 .167 5.32 .164 5.36 .162
       5.40 .159 5.44 .157 5.48 .155 5.52 .152 5.56 .150
       5.60 .148 5.64 .146 5.68 .144 5.72 .142 5.76 .139
       5.80 .137 5.84 .135 5.88 .133 5.92 .131 5.96 .129
       6.00 .127 /
\setdashes <4pt>
\plot
   0 1.000 .04 1.000 .08 .999 .12 .997 .16 .994
       .20 .991 .24 .987 .28 .982 .32 .977 .36 .971
       .40 .964 .44 .956 .48 .947 .52 .938 .56 .928
       .60 .917 .64 .906 .68 .893 .72 .880 .76 .866
       .80 .851 .84 .835 .88 .819 .92 .801 .96 .783
       1.00 .764 1.04 .744 1.08 .725 1.12 .706 1.16 .688
       1.20 .671 1.24 .654 1.28 .637 1.32 .621 1.36 .605
       1.40 .589 1.44 .574 1.48 .560 1.52 .545 1.56 .531
       1.60 .518 1.64 .504 1.68 .492 1.72 .479 1.76 .467
       1.80 .455 1.84 .443 1.88 .432 1.92 .421 1.96 .410
       2.00 .400 2.04 .389 2.08 .379 2.12 .370 2.16 .360
       2.20 .351 2.24 .342 2.28 .333 2.32 .325 2.36 .316
       2.40 .308 2.44 .300 2.48 .293 2.52 .285 2.56 .278
       2.60 .271 2.64 .264 2.68 .257 2.72 .251 2.76 .244
       2.80 .238 2.84 .232 2.88 .226 2.92 .220 2.96 .215
       3.00 .209 3.04 .204 3.08 .198 3.12 .193 3.16 .188
       3.20 .184 3.24 .179 3.28 .174 3.32 .170 3.36 .166
       3.40 .161 3.44 .157 3.48 .153 3.52 .149 3.56 .145
       3.60 .142 3.64 .138 3.68 .135 3.72 .131 3.76 .128
       3.80 .125 3.84 .121 3.88 .118 3.92 .115 3.96 .112
       4.00 .109 4.04 .107 4.08 .104 4.12 .101 4.16 .099
       4.20 .096 4.24 .094 4.28 .091 4.32 .089 4.36 .087
       4.40 .084 4.44 .082 4.48 .080 4.52 .078 4.56 .076
       4.60 .074 4.64 .072 4.68 .070 4.72 .069 4.76 .067
       4.80 .065 4.84 .063 4.88 .062 4.92 .060 4.96 .059
       5.00 .057 5.04 .056 5.08 .054 5.12 .053 5.16 .052
       5.20 .050 5.24 .049 5.28 .048 5.32 .047 5.36 .045
       5.40 .044 5.44 .043 5.48 .042 5.52 .041 5.56 .040
       5.60 .039 5.64 .038 5.68 .037 5.72 .036 5.76 .035
       5.80 .034 5.84 .033 5.88 .032 5.92 .032 5.96 .031
       6.00 .030 /
\setdashes <2pt>
\plot
   0 1.000 .04 1.000 .08 .999 .12 .997 .16 .995
       .20 .992 .24 .988 .28 .984 .32 .980 .36 .974
       .40 .968 .44 .962 .48 .955 .52 .947 .56 .939
       .60 .931 .64 .922 .68 .912 .72 .902 .76 .892
       .80 .881 .84 .870 .88 .859 .92 .847 .96 .835
       1.00 .823 1.04 .811 1.08 .800 1.12 .788 1.16 .776
       1.20 .765 1.24 .754 1.28 .743 1.32 .732 1.36 .721
       1.40 .711 1.44 .701 1.48 .690 1.52 .680 1.56 .670
       1.60 .661 1.64 .651 1.68 .641 1.72 .632 1.76 .623
       1.80 .614 1.84 .605 1.88 .596 1.92 .587 1.96 .579
       2.00 .570 2.04 .562 2.08 .554 2.12 .546 2.16 .538
       2.20 .530 2.24 .522 2.28 .515 2.32 .507 2.36 .500
       2.40 .492 2.44 .485 2.48 .478 2.52 .471 2.56 .464
       2.60 .458 2.64 .451 2.68 .444 2.72 .438 2.76 .431
       2.80 .425 2.84 .419 2.88 .413 2.92 .407 2.96 .401
       3.00 .395 3.04 .389 3.08 .384 3.12 .378 3.16 .373
       3.20 .367 3.24 .362 3.28 .356 3.32 .351 3.36 .346
       3.40 .341 3.44 .336 3.48 .331 3.52 .326 3.56 .322
       3.60 .317 3.64 .312 3.68 .308 3.72 .303 3.76 .299
       3.80 .294 3.84 .290 3.88 .286 3.92 .282 3.96 .278
       4.00 .274 4.04 .270 4.08 .266 4.12 .262 4.16 .258
       4.20 .254 4.24 .251 4.28 .247 4.32 .243 4.36 .240
       4.40 .236 4.44 .233 4.48 .229 4.52 .226 4.56 .223
       4.60 .220 4.64 .216 4.68 .213 4.72 .210 4.76 .207
       4.80 .204 4.84 .201 4.88 .198 4.92 .195 4.96 .192
       5.00 .190 5.04 .187 5.08 .184 5.12 .181 5.16 .179
       5.20 .176 5.24 .174 5.28 .171 5.32 .169 5.36 .166
       5.40 .164 5.44 .161 5.48 .159 5.52 .157 5.56 .154
       5.60 .152 5.64 .150 5.68 .148 5.72 .145 5.76 .143
       5.80 .141 5.84 .139 5.88 .137 5.92 .135 5.96 .133
       6.00 .131 /
\setsolid
\plot
0 1.000 .04 1.000 .08 .999 .12 .997 .16 .994
       .20 .991 .24 .987 .28 .983 .32 .978 .36 .972
       .40 .965 .44 .958 .48 .950 .52 .942 .56 .932
       .60 .922 .64 .912 .68 .901 .72 .889 .76 .877
       .80 .864 .84 .850 .88 .836 .92 .821 .96 .806
       1.00 .790 1.04 .774 1.08 .758 1.12 .742 1.16 .727
       1.20 .713 1.24 .698 1.28 .684 1.32 .670 1.36 .656
       1.40 .643 1.44 .630 1.48 .617 1.52 .605 1.56 .592
       1.60 .580 1.64 .568 1.68 .557 1.72 .546 1.76 .535
       1.80 .524 1.84 .513 1.88 .503 1.92 .492 1.96 .482
       2.00 .473 2.04 .463 2.08 .454 2.12 .444 2.16 .435
       2.20 .426 2.24 .418 2.28 .409 2.32 .401 2.36 .393
       2.40 .385 2.44 .377 2.48 .369 2.52 .362 2.56 .354
       2.60 .347 2.64 .340 2.68 .333 2.72 .327 2.76 .320
       2.80 .313 2.84 .307 2.88 .301 2.92 .295 2.96 .289
       3.00 .283 3.04 .277 3.08 .271 3.12 .266 3.16 .261
       3.20 .255 3.24 .250 3.28 .245 3.32 .240 3.36 .235
       3.40 .230 3.44 .226 3.48 .221 3.52 .217 3.56 .212
       3.60 .208 3.64 .204 3.68 .199 3.72 .195 3.76 .191
       3.80 .188 3.84 .184 3.88 .180 3.92 .176 3.96 .173
       4.00 .169 4.04 .166 4.08 .162 4.12 .159 4.16 .156
       4.20 .153 4.24 .150 4.28 .147 4.32 .144 4.36 .141
       4.40 .138 4.44 .135 4.48 .132 4.52 .130 4.56 .127
       4.60 .124 4.64 .122 4.68 .119 4.72 .117 4.76 .115
       4.80 .112 4.84 .110 4.88 .108 4.92 .106 4.96 .103
       5.00 .101 5.04 .099 5.08 .097 5.12 .095 5.16 .093
       5.20 .091 5.24 .090 5.28 .088 5.32 .086 5.36 .084
       5.40 .082 5.44 .081 5.48 .079 5.52 .078 5.56 .076
       5.60 .074 5.64 .073 5.68 .071 5.72 .070 5.76 .069
       5.80 .067 5.84 .066 5.88 .064 5.92 .063 5.96 .062
       6.00 .061 /
\endpicture$$
\caption{\label{fig-3} $2mV_0a^2=0.7$}
\end{figure}}

{\figsize\begin{figure}[f]
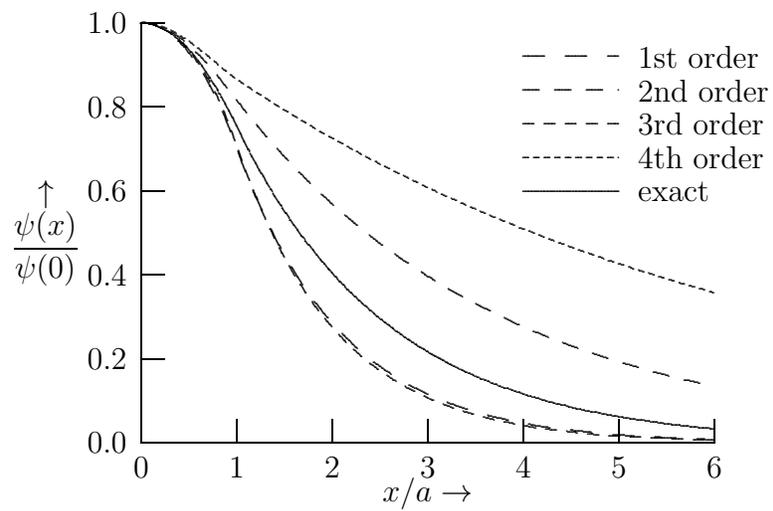

$$\beginpicture
\setcoordinatesystem units <.5truein,2.2truein>
\setplotarea x from 0 to 6, y from 0 to 1
\axis bottom visible ticks in numbered from 0 to 6 by 1 /
\axis left visible ticks in numbered from 0.0 to 1.0 by .2 /
\put {$\displaystyle \uparrow\atop\displaystyle
{\psi(x)\over\psi(0)}$} at -1 .5
\put {$x/a \rightarrow$} at 3 -.12
\setdashes <8pt>
\plot 4 .92 5 .92 /
\put {1st order} [l] at 5.2 .92
\setdashes <6pt>
\plot 4 .84 5 .84 /
\put {2nd order} [l] at 5.2 .84
\setdashes <4pt>
\plot 4 .76 5 .76 /
\put {3rd order} [l] at 5.2 .76
\setdashes <2pt>
\plot 4 .68 5 .68 /
\put {4th order} [l] at 5.2 .68
\setsolid
\plot 4 .6 5 .6 /
\put {exact} [l] at 5.2 .6
\setdashes <8pt>
\plot
   0 1.000 .04 1.000 .08 .998 .12 .996 .16 .993
       .20 .989 .24 .984 .28 .978 .32 .971 .36 .964
       .40 .955 .44 .946 .48 .935 .52 .924 .56 .911
       .60 .898 .64 .883 .68 .868 .72 .851 .76 .833
       .80 .815 .84 .795 .88 .774 .92 .751 .96 .728
       1.00 .703 1.04 .678 1.08 .654 1.12 .631 1.16 .609
       1.20 .587 1.24 .567 1.28 .547 1.32 .527 1.36 .509
       1.40 .491 1.44 .473 1.48 .457 1.52 .440 1.56 .425
       1.60 .410 1.64 .395 1.68 .381 1.72 .368 1.76 .355
       1.80 .342 1.84 .330 1.88 .319 1.92 .307 1.96 .296
       2.00 .286 2.04 .276 2.08 .266 2.12 .257 2.16 .248
       2.20 .239 2.24 .230 2.28 .222 2.32 .214 2.36 .207
       2.40 .199 2.44 .192 2.48 .186 2.52 .179 2.56 .173
       2.60 .167 2.64 .161 2.68 .155 2.72 .150 2.76 .144
       2.80 .139 2.84 .134 2.88 .130 2.92 .125 2.96 .121
       3.00 .116 3.04 .112 3.08 .108 3.12 .104 3.16 .101
       3.20 .097 3.24 .094 3.28 .090 3.32 .087 3.36 .084
       3.40 .081 3.44 .078 3.48 .075 3.52 .073 3.56 .070
       3.60 .068 3.64 .065 3.68 .063 3.72 .061 3.76 .059
       3.80 .057 3.84 .055 3.88 .053 3.92 .051 3.96 .049
       4.00 .047 4.04 .046 4.08 .044 4.12 .042 4.16 .041
       4.20 .039 4.24 .038 4.28 .037 4.32 .035 4.36 .034
       4.40 .033 4.44 .032 4.48 .031 4.52 .030 4.56 .029
       4.60 .028 4.64 .027 4.68 .026 4.72 .025 4.76 .024
       4.80 .023 4.84 .022 4.88 .021 4.92 .021 4.96 .020
       5.00 .019 5.04 .019 5.08 .018 5.12 .017 5.16 .017
       5.20 .016 5.24 .015 5.28 .015 5.32 .014 5.36 .014
       5.40 .013 5.44 .013 5.48 .012 5.52 .012 5.56 .012
       5.60 .011 5.64 .011 5.68 .010 5.72 .010 5.76 .010
       5.80 .009 5.84 .009 5.88 .009 5.92 .008 5.96 .008
       6.00 .008 /
\setdashes <6pt>
\plot
   0 1.000 .04 1.000 .08 .999 .12 .997 .16 .994
       .20 .991 .24 .987 .28 .983 .32 .978 .36 .972
       .40 .966 .44 .959 .48 .951 .52 .943 .56 .934
       .60 .925 .64 .916 .68 .906 .72 .896 .76 .885
       .80 .874 .84 .863 .88 .851 .92 .840 .96 .828
       1.00 .816 1.04 .805 1.08 .793 1.12 .782 1.16 .771
       1.20 .760 1.24 .749 1.28 .738 1.32 .727 1.36 .717
       1.40 .707 1.44 .697 1.48 .687 1.52 .677 1.56 .667
       1.60 .658 1.64 .648 1.68 .639 1.72 .630 1.76 .621
       1.80 .612 1.84 .603 1.88 .595 1.92 .586 1.96 .578
       2.00 .569 2.04 .561 2.08 .553 2.12 .545 2.16 .538
       2.20 .530 2.24 .522 2.28 .515 2.32 .508 2.36 .500
       2.40 .493 2.44 .486 2.48 .479 2.52 .472 2.56 .466
       2.60 .459 2.64 .452 2.68 .446 2.72 .439 2.76 .433
       2.80 .427 2.84 .421 2.88 .415 2.92 .409 2.96 .403
       3.00 .397 3.04 .392 3.08 .386 3.12 .381 3.16 .375
       3.20 .370 3.24 .364 3.28 .359 3.32 .354 3.36 .349
       3.40 .344 3.44 .339 3.48 .334 3.52 .329 3.56 .325
       3.60 .320 3.64 .316 3.68 .311 3.72 .307 3.76 .302
       3.80 .298 3.84 .294 3.88 .289 3.92 .285 3.96 .281
       4.00 .277 4.04 .273 4.08 .269 4.12 .265 4.16 .262
       4.20 .258 4.24 .254 4.28 .251 4.32 .247 4.36 .243
       4.40 .240 4.44 .237 4.48 .233 4.52 .230 4.56 .227
       4.60 .223 4.64 .220 4.68 .217 4.72 .214 4.76 .211
       4.80 .208 4.84 .205 4.88 .202 4.92 .199 4.96 .196
       5.00 .193 5.04 .191 5.08 .188 5.12 .185 5.16 .183
       5.20 .180 5.24 .177 5.28 .175 5.32 .172 5.36 .170
       5.40 .167 5.44 .165 5.48 .163 5.52 .160 5.56 .158
       5.60 .156 5.64 .154 5.68 .151 5.72 .149 5.76 .147
       5.80 .145 5.84 .143 5.88 .141 5.92 .139 5.96 .137
       6.00 .135 /
\setdashes <4pt>
\plot
   0 1.000 .04 1.000 .08 .998 .12 .996 .16 .994
       .20 .990 .24 .985 .28 .980 .32 .974 .36 .967
       .40 .959 .44 .950 .48 .940 .52 .929 .56 .917
       .60 .904 .64 .891 .68 .875 .72 .859 .76 .842
       .80 .823 .84 .803 .88 .782 .92 .759 .96 .734
       1.00 .709 1.04 .682 1.08 .657 1.12 .633 1.16 .609
       1.20 .587 1.24 .565 1.28 .544 1.32 .524 1.36 .505
       1.40 .486 1.44 .468 1.48 .451 1.52 .434 1.56 .418
       1.60 .402 1.64 .387 1.68 .373 1.72 .359 1.76 .346
       1.80 .333 1.84 .321 1.88 .309 1.92 .298 1.96 .286
       2.00 .276 2.04 .266 2.08 .256 2.12 .246 2.16 .237
       2.20 .228 2.24 .220 2.28 .212 2.32 .204 2.36 .196
       2.40 .189 2.44 .182 2.48 .175 2.52 .169 2.56 .163
       2.60 .157 2.64 .151 2.68 .145 2.72 .140 2.76 .135
       2.80 .130 2.84 .125 2.88 .120 2.92 .116 2.96 .112
       3.00 .107 3.04 .103 3.08 .100 3.12 .096 3.16 .092
       3.20 .089 3.24 .086 3.28 .082 3.32 .079 3.36 .076
       3.40 .074 3.44 .071 3.48 .068 3.52 .066 3.56 .063
       3.60 .061 3.64 .059 3.68 .057 3.72 .054 3.76 .052
       3.80 .051 3.84 .049 3.88 .047 3.92 .045 3.96 .043
       4.00 .042 4.04 .040 4.08 .039 4.12 .037 4.16 .036
       4.20 .035 4.24 .033 4.28 .032 4.32 .031 4.36 .030
       4.40 .029 4.44 .028 4.48 .027 4.52 .026 4.56 .025
       4.60 .024 4.64 .023 4.68 .022 4.72 .021 4.76 .020
       4.80 .020 4.84 .019 4.88 .018 4.92 .018 4.96 .017
       5.00 .016 5.04 .016 5.08 .015 5.12 .015 5.16 .014
       5.20 .013 5.24 .013 5.28 .013 5.32 .012 5.36 .012
       5.40 .011 5.44 .011 5.48 .010 5.52 .010 5.56 .010
       5.60 .009 5.64 .009 5.68 .009 5.72 .008 5.76 .008
       5.80 .008 5.84 .007 5.88 .007 5.92 .007 5.96 .007
       6.00 .006 /
\setdashes <2pt>
\plot
   0 1.000 .04 1.001 .08 .998 .12 .998 .16 .996
       .20 .993 .24 .990 .28 .986 .32 .982 .36 .977
       .40 .972 .44 .965 .48 .959 .52 .954 .56 .946
       .60 .940 .64 .931 .68 .925 .72 .917 .76 .910
       .80 .902 .84 .894 .88 .887 .92 .880 .96 .873
       1.00 .866 1.04 .860 1.08 .854 1.12 .848 1.16 .842
       1.20 .836 1.24 .830 1.28 .824 1.32 .819 1.36 .813
       1.40 .807 1.44 .801 1.48 .796 1.52 .790 1.56 .785
       1.60 .779 1.64 .774 1.68 .768 1.72 .763 1.76 .757
       1.80 .752 1.84 .747 1.88 .741 1.92 .736 1.96 .731
       2.00 .726 2.04 .721 2.08 .716 2.12 .711 2.16 .706
       2.20 .701 2.24 .696 2.28 .691 2.32 .686 2.36 .681
       2.40 .676 2.44 .672 2.48 .667 2.52 .662 2.56 .657
       2.60 .653 2.64 .648 2.68 .644 2.72 .639 2.76 .635
       2.80 .630 2.84 .626 2.88 .621 2.92 .617 2.96 .613
       3.00 .608 3.04 .604 3.08 .600 3.12 .595 3.16 .591
       3.20 .587 3.24 .583 3.28 .579 3.32 .575 3.36 .571
       3.40 .567 3.44 .563 3.48 .559 3.52 .555 3.56 .551
       3.60 .547 3.64 .543 3.68 .539 3.72 .536 3.76 .532
       3.80 .528 3.84 .524 3.88 .521 3.92 .517 3.96 .513
       4.00 .510 4.04 .506 4.08 .503 4.12 .499 4.16 .496
       4.20 .492 4.24 .489 4.28 .485 4.32 .482 4.36 .478
       4.40 .475 4.44 .472 4.48 .468 4.52 .465 4.56 .462
       4.60 .458 4.64 .455 4.68 .452 4.72 .449 4.76 .446
       4.80 .443 4.84 .439 4.88 .436 4.92 .433 4.96 .430
       5.00 .427 5.04 .424 5.08 .421 5.12 .418 5.16 .415
       5.20 .412 5.24 .409 5.28 .407 5.32 .404 5.36 .401
       5.40 .398 5.44 .395 5.48 .392 5.52 .390 5.56 .387
       5.60 .384 5.64 .381 5.68 .379 5.72 .376 5.76 .373
       5.80 .371 5.84 .368 5.88 .366 5.92 .363 5.96 .361
       6.00 .358 /
\setsolid
\plot
   0 1.000 .04 1.000 .08 .998 .12 .996 .16 .993
       .20 .990 .24 .985 .28 .980 .32 .974 .36 .967
       .40 .959 .44 .951 .48 .942 .52 .932 .56 .921
       .60 .909 .64 .897 .68 .884 .72 .870 .76 .856
       .80 .840 .84 .825 .88 .808 .92 .791 .96 .773
       1.00 .755 1.04 .736 1.08 .718 1.12 .700 1.16 .683
       1.20 .666 1.24 .650 1.28 .634 1.32 .618 1.36 .603
       1.40 .588 1.44 .574 1.48 .560 1.52 .546 1.56 .532
       1.60 .519 1.64 .507 1.68 .494 1.72 .482 1.76 .470
       1.80 .459 1.84 .447 1.88 .436 1.92 .426 1.96 .415
       2.00 .405 2.04 .395 2.08 .385 2.12 .376 2.16 .366
       2.20 .357 2.24 .349 2.28 .340 2.32 .332 2.36 .324
       2.40 .316 2.44 .308 2.48 .300 2.52 .293 2.56 .286
       2.60 .279 2.64 .272 2.68 .265 2.72 .259 2.76 .252
       2.80 .246 2.84 .240 2.88 .234 2.92 .228 2.96 .223
       3.00 .217 3.04 .212 3.08 .207 3.12 .202 3.16 .197
       3.20 .192 3.24 .187 3.28 .182 3.32 .178 3.36 .174
       3.40 .169 3.44 .165 3.48 .161 3.52 .157 3.56 .153
       3.60 .150 3.64 .146 3.68 .142 3.72 .139 3.76 .135
       3.80 .132 3.84 .129 3.88 .126 3.92 .123 3.96 .120
       4.00 .117 4.04 .114 4.08 .111 4.12 .108 4.16 .106
       4.20 .103 4.24 .100 4.28 .098 4.32 .096 4.36 .093
       4.40 .091 4.44 .089 4.48 .086 4.52 .084 4.56 .082
       4.60 .080 4.64 .078 4.68 .076 4.72 .074 4.76 .073
       4.80 .071 4.84 .069 4.88 .067 4.92 .066 4.96 .064
       5.00 .063 5.04 .061 5.08 .060 5.12 .058 5.16 .057
       5.20 .055 5.24 .054 5.28 .053 5.32 .051 5.36 .050
       5.40 .049 5.44 .048 5.48 .046 5.52 .045 5.56 .044
       5.60 .043 5.64 .042 5.68 .041 5.72 .040 5.76 .039
       5.80 .038 5.84 .037 5.88 .036 5.92 .035 5.96 .034
       6.00 .034 /
\endpicture$$
\caption{\label{fig-4}
{$2mV_0a^2=0.9$}
}\end{figure}}

\section*{Acknowledgements} We are very grateful to David Kaplan for many
interesting conversations.

\appendix
\section{Appendix:  Expansion Formulae}
This appendix lists the energy and $\chi$ expansions to
several orders in $mV$; the structure of the collection of terms
present at a particular order in the $\kappa a$ expansion has been
described in section 2.

\begin{equation}
\begin{array}{c}
\kappa  =  -m\int dx\,V(x) - m^2\int dx_1 \,dx_2\, |x_1-x_2|\,
V(x_1)V(x_2)\\
  -m^3\Biggl[
\int dx_1\,dx_2\,dx_3\, |x_1-x_2| |x_2-x_3|\, V(x_1)V(x_2)V(x_3) \\
  + {1\over2} \Biggl(\int dx\, V(x)\Biggr)
\Biggl(\int dx_1\,dx_2\, |x_1-x_2|^2 V(x_1)V(x_2)\Biggr) \Biggr] \\
  - m^4 \Biggl[
\int dx_1\,dx_2\,dx_3\,dx_4\, |x_1-x_2| |x_2-x_3| |x_3-x_4|
\,V(x_1)V(x_2)V(x_3)V(x_4) \\
  + \Biggl(\int dx\, V(x)\Biggr) \Biggl(
\int dx_1\,dx_2\,dx_3\, |x_1-x_2|^2 |x_2-x_3|\, V(x_1)V(x_2)V(x_3) \Biggr) \\
  + {1\over 2} \Biggl(\int dx_1\,dx_2\, |x_1-x_2|^2 V(x_1)\,V(x_2)\Biggr)
\Biggl(\int dx_1\,dx_2\, |x_1-x_2|\, V(x_1)\,V(x_2)\Biggr) \\
  + {1\over 6}\Biggl(\int dx\, V(x)\Biggr)^2
\Biggl(\int dx_1\,dx_2\, |x_1-x_2|^3 V(x_1)V(x_2)\Biggr) \Biggr] \\
  + {\cal O}\bigl(m^5V^5\bigr)
\end{array}
\label{kappa}
\end{equation}

\begin{equation}
\begin{array}{c}
-\kappa\chi(x_1) = m \, V(x_1)
+ m^2 \int dx_2 |x_1-x_2|\, V(x_1)V(x_2) \\
 + m^3 \Biggl[
\int dx_2\,dx_3 |x_1-x_2||x_2-x_3|\, V(x_1)V(x_2)V(x_3) \\
 +{1\over2}
\Biggl( \int dx_2\, |x_1-x_2|^2 V(x_1)V(x_2) \Biggr)
\Biggl( \int dx\,V(x) \Biggr)
\Biggr] \\
 + m^4 \Biggl[
\int dx_2\,dx_3\,dx_4 |x_1-x_2||x_2-x_3||x_3-x_4|\,
V(x_1)\,V(x_2)\,V(x_3)\,V(x_4) \\
 + {1\over 2}
\Biggl( \int dx_2\,dx_3\, |x_1-x_2||x_2-x_3|^2
V(x_1)V(x_2)V(x_3) \Biggr)
\Biggl( \int dx\, V(x) \Biggr) \\
 + {1\over 2}
\Biggl( \int dx_2\,dx_3\, |x_1-x_2|^2|x_2-x_3|\,
V(x_1)V(x_2)V(x_3) \Biggr)
\Biggl( \int dx\, V(x) \Biggr) \\
 +{1\over 2}
\Biggl( \int dx_2\, |x_1-x_2|^2 V(x_1)V(x_2) \Biggr)
\Biggl( \int dx_2\, |x_1-x_2|\, V(x_1)V(x_2) \Biggr) \\
 +{1\over 6}
\Biggl( \int dx_2\, |x_1-x_2|^3 V(x_1)V(x_2) \Biggr)
\Biggl( \int dx\, V(x) \Biggr)^2 \Biggr] \\
 + {\cal O}\bigl(m^5V^5\bigr)
\end{array}
\label{chia}
\end{equation}

\end{document}